# DeltaHedge: A Multi-Agent Framework for Portfolio Options Optimization

*Completed Research Paper*

**Feliks Bańka**
Warsaw University of Technology
Warsaw, Poland
feliks.banka.stud@pw.edu.pl

**Jarosław A. Chudziak**
Warsaw University of Technology
Warsaw, Poland
jaroslaw.chudziak@pw.edu.pl

## Abstract

*In volatile financial markets, balancing risk and return remains a significant challenge. Traditional approaches often focus solely on equity allocation, overlooking the strategic advantages of options trading for dynamic risk hedging. This work presents **DeltaHedge**, a multi-agent framework that integrates options trading with AI-driven portfolio management. By combining advanced reinforcement learning techniques with an ensembled options-based hedging strategy, DeltaHedge enhances risk-adjusted returns and stabilizes portfolio performance across varying market conditions. Experimental results demonstrate that DeltaHedge outperforms traditional strategies and standalone models, underscoring its potential to transform practical portfolio management in complex financial environments. Building on these findings, this paper contributes to the fields of quantitative finance and AI-driven portfolio optimization by introducing a novel multi-agent system for integrating options trading strategies, addressing a gap in the existing literature.*

**Keywords:** Artificial Intelligence, Machine Learning, Multi-Agent Systems, Portfolio Management, Options Trading

## Introduction

Effective portfolio management is essential in contemporary finance to maintain a balance between risk and return in volatile markets. Options stand out among financial instruments for their dual roles in risk hedging and speculative opportunities (Black, 1975; Hull, 2006). Despite their potential, the integration of options trading into AI-driven portfolio management systems remains underexplored. Existing frameworks largely focus on equity and asset allocation strategies, overlooking the unique challenges and opportunities that options trading presents (Bilokon, 2023; Hans Buehler Lukas Gonon & Wood, 2019).

Recent advancements in machine learning (ML) and multi-agent systems (MAS) have enabled sophisticated approaches to financial decision-making (Wawer et al., 2024). Techniques such as Deep Reinforcement Learning (DRL) (Y. Liu et al., 2024; Sarin et al., 2024) have demonstrated efficacy in forecasting and trading. MAS frameworks, employing specialized agents for market prediction and strategy execution, have become popular in finance (Huang & Tanaka, 2022; Lussange et al., 2021). Moreover, ensemble approaches that combine multiple reinforcement learning (RL) agents have shown promise in reducing individual model biases and adjusting to changing market conditions (Yang et al., 2020). However, these techniques rarely incorporate options trading and hedging strategies.

To address this gap, we propose a DeltaHedge, a multi-agent framework that integrates options trading into AI-driven portfolio management. Unlike conventional equity-focused methods, DeltaHedge coordinates specialized modules for forecasting, sentiment analysis, trading, and dynamic hedging. By coordinating these agents under a central mechanism, our system enhances risk-adjusted returns and demonstrates how options can be effectively leveraged in a multi-agent setting to manage complex market conditions.

This paper makes several notable contributions to the field of AI-driven portfolio management and multi-agent finance. First, it introduces DeltaHedge, a novel multi-agent framework that integrates options-based hedging into traditional equity trading workflows. By unifying specialized agents, our approach addresses the limitations of conventional asset allocation methods and expands the applicability of AI-driven decision-making in volatile markets.

Second, our research proposes an ensemble-based hedging strategy designed to mitigate algorithmic biases and adapt effectively to evolving market conditions. Rather than relying on a single reinforcement learning model, our approach dynamically selects among multiple RL agents in real time. This mechanism offers deeper flexibility and robustness when market regimes shift, demonstrating that ensemble techniques can improve both upside potential and downside protection. These insights into ensemble RL methodology have broader implications for financial applications that require consistent performance in the face of rapidly changing data distributions.

Third, we present a comprehensive benchmarking and ablation study that evaluates DeltaHedge against traditional trading methods, standalone RL baselines, state-of-the-art frameworks, and systems without hedging. The results reveal how each component—particularly the integration of options-based hedging and the ensemble mechanism—contributes to significant gains in risk-adjusted returns and portfolio stability. By providing a transparent analysis of these components, our work serves as a foundation for researchers to explore which facets of multi-agent and ensemble learning are most impactful under different market scenarios.

The remainder of the paper is structured as follows. In Section 2, we provide an overview of related work. Section 3 details our multi-agent framework, and Section 4 describes the experimental setup and presents the results. Finally, Section 5 concludes with remarks on the findings and offers avenues for future exploration.

# Related work

Portfolio management has been a central focus in quantitative finance, with significant advancements in optimization techniques and the use of options for risk reduction and return improvement. This section reviews existing literature on portfolio management and options, multi-agent systems in finance, and machine learning models for financial applications.

## *Portfolio Management and Options*

Options trading has long been recognized as a powerful tool for managing risk and enhancing portfolio returns (Hull, 2006). Foundational works such as Black and Scholes' option pricing model (Black, 1975) and Merton's rational option pricing theory (Merton, 1973) established the theoretical underpinnings for options valuation. These models were later extended by Cox, Ross, and Rubinstein's binomial option pricing approach, which provided a more intuitive framework for pricing derivatives (Cox et al., 1979).

Subsequent research has explored the integration of options into portfolio optimization (Andersson & Oosterlee, 2023; Bańka & Chudziak, 2025b; Chavas et al., 2024; Pang et al., 2022; Turan G. Bali Heiner Beckmeyer & Weigert, 2023). Casas and Veiga investigated the role of volatility asymmetry in option pricing and hedging (Casas & Veiga, 2021). More recently, Sun et al. combined transformer-based reinforcement learning with the Black-Litterman model for portfolio optimization, emphasizing the role of options in balancing risk and return (Sun et al., 2024).

Despite these advancements, the application of options in AI-driven portfolio management remains limited (Turan G. Bali Heiner Beckmeyer & Weigert, 2023). Many existing frameworks focus on equity allocation, neglecting the nuanced benefits of options trading (Tripathi & Bhavanani, 2024). For instance, Huang and Tanaka proposed a multi-agent reinforcement learning system for portfolio management but did not incorporate options (Huang & Tanaka, 2022). Similarly, trading systems like DeepTrader prioritize asset allocation strategies without addressing the complexities of options hedging (Wang et al., 2021). Equity-only AI systems remain exposed to discontinuous price jumps and are unable to harvest the volatility-risk premium that options offer. Pang et al. show that incorporating near-ATM protective puts into portfolio optimization materially reduces tail-risk and drawdowns under realistic frictions (Pang et al., 2022), while Escobar-Anel et al. quantify significant expected-utility gains from such volatility-based hedging strategies (Escobar-Anel et al., 2022). This gap highlights the need for frameworks that systematically integrate options into portfolio decision-making.

### *Multi-Agent Systems in Finance*

MAS have been increasingly adopted in finance to automate decision-making processes across forecasting, trading, and risk management. Early work by Lussange et al. introduced multi-agent reinforcement-learning architectures for stock-market modeling, emphasizing the benefits of distributed decision-making in financial markets (Lussange et al., 2021). Modern MAS frameworks have further advanced these ideas, leveraging reinforcement learning and explainable AI techniques (Lee et al., 2020).

For example, Li et al. proposed TradingGPT, a multi-agent system with layered memory and distinct agent roles for enhanced trading performance (Y. Li et al., 2023). Similarly, Yu et al. introduced a hierarchical reinforcement learning approach for multi-agent coordination, demonstrating its effectiveness in target-oriented financial tasks (Yu et al., 2024). FinAgent provides another recent example of a modular multi-agent architecture, integrating forecasting, asset management, and order execution for equity portfolios (Zhang et al., 2024). These systems highlight the potential of MAS for managing complex interactions in trading and portfolio optimization (Guo et al., 2024).

However, most MAS implementations in finance focus on equities and do not address the integration of options trading. Belcak et al. developed a simulation framework for agent-based trading but limit its scope to latency effects in financial markets (Belcak et al., 2022). Additionally, systems like SARIN's multi-agent reinforcement learning framework (Sarin et al., 2024) and MSPM (Huang & Tanaka, 2022) prioritize equities, leaving a significant gap in the application of MAS to options-based strategies.

### *Machine Learning in Financial Modeling*

Recent progress in ML has greatly influenced quantitative finance, particularly in time-series forecasting and sentiment analysis—two crucial drivers of trading decisions (Lim & Zohren, 2021; Tan et al., 2024). While early deep-learning approaches relied on recurrent networks such as LSTM (Hochreiter & Schmidhuber, 1997) to capture sequential dependencies, transformer-based architectures like Informer (Bańka & Chudziak, 2025a; Wen et al., 2023; Zhou et al., 2021) have more recently demonstrated superior ability to model long-range patterns and greater computational efficiency. At the same time, sentiment tools (e.g., FinBERT) have proven effective in extracting market sentiment from financial news and social media (Araci, 2019). Integrating these signals into trading systems can improve adaptability and predictive accuracy (Y. Liu et al., 2024).

Beyond supervised methods, RL has gained traction for its ability to frame trading and hedging as sequential decision processes. Although ensemble RL strategies have shown promise in mitigating individual model biases (Yang et al., 2020), these techniques have seen limited deployment in multi-agent contexts that specifically address options trading. Consequently, there remains a strong motivation to develop comprehensive frameworks that integrate advanced RL, multi-agent architectures, and the distinct benefits of options hedging.

## DeltaHedge Framework

This section presents **DeltaHedge**—a hierarchical multi-agent system for portfolio optimization, integrating predictive modeling, sentiment analysis, and dynamic options-based hedging. Reinforcement learning facilitates adaptive decision-making within the Trading and Hedging Agents, with the Trading Agent specifically employing Proximal Policy Optimization (PPO) (Hans Buehler Lukas Gonon & Wood, 2019; Sharma et al., 2024). DeltaHedge incorporates a transformer-based forecasting module (Informer), which provides long-range temporal structure and short-term price movement predictions (Huang & Tanaka, 2022; Y. Liu et al., 2024; Sarin et al., 2024). Sentiment analysis offers quantitative insights for risk management by incorporating market stance indicators derived from sentiment trend movements and volatility sensitivity (Araci, 2019; Guo et al., 2024). The management of this process is overseen by the Coordinator Agent, as illustrated in Figure 1.

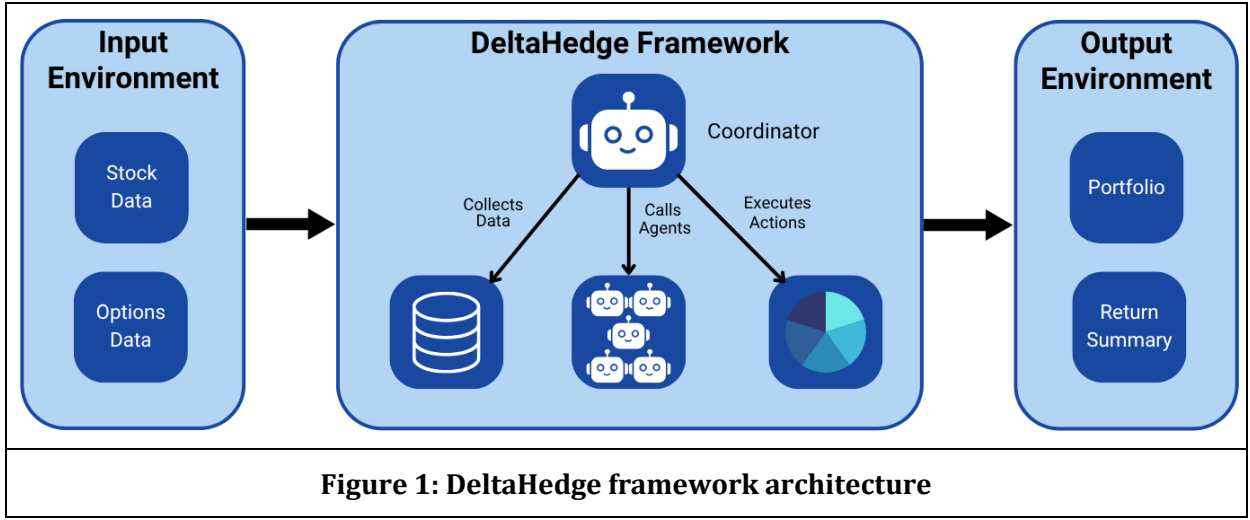

**Figure 1: DeltaHedge framework architecture**

### *Data Overview and Assumptions*

The proposed framework relies on several key data sources:

- **Daily Price Data**: Includes open, high, low, close, and volume (OHLCV) metrics for the underlying asset.

- **Protective Options**: Near-the-money options with a fixed time-to-expiration (e.g., 30 days), chosen based on liquidity and pricing stability.

- **Sentiment Data**: Daily sentiment scores computed from financial news headlines using a transformer-based sentiment classification model.

- **Volatility Index**: An external measure (VIX indicator) representing broader market uncertainty.

The state at time t is defined as:

$$s_t = [p_t, h_t, b_t, n_t, O_t, f_t, sent_t, VIX_t]$$

where: $p_t$ is the current asset price, $h_t$ is the number of shares held, $b_t$ is the cash balance, $n_t$ is the number of protective put options held, $O_t$ represents an option contract, $f_t$ is the forecasted percentage change from the Forecasting Agent, $sent_t$ is the sentiment score, and $VIX_t$ is the volatility index. The portfolio value is given then by: $V_t = b_t + p_t h_t + O_{price_t} n_t$.

We use daily data since most market crashes and large return jumps occur overnight, outside intraday windows. Additionally, our use of 30-day protective puts makes daily rebalancing sufficient, while avoiding excessive churn and transaction costs. The system exclusively engages in long positions, excluding short selling. This decision aligns with theoretical considerations emphasizing hedging as a mechanism for downside risk mitigation rather than speculative trading. Protective options are incorporated strictly to preserve risk-adjusted returns.

### *Specialized Agents*

The framework consists of multiple specialized agents that interact dynamically to optimize both stock trading and options-based hedging based on predictive signals and sentiment-driven risk assessment. The Forecasting and Sentiment Analysis Agents generate market signals that directly inform the decision-making of the Trading and Hedging Agents. The Coordinator Agent ensures that the actions taken are consistent with portfolio constraints and risk policies.

**Forecasting Agent**

The Forecasting Agent utilizes Informer, a transformer-based sequence model with probabilistic sparse self-attention (Zhou et al., 2021), to predict short-term asset price movements. It processes historical OHLCV data from the past 60 days and outputs a percentage change forecast for the next 30 days (e.g., +2%), which is used by downstream agents to adjust their positions defined as $f_{t+1:t+30} = \text{Informer}(p_{t-60:t})$.

**Sentiment Analysis Agent**

The Sentiment Analysis Agent uses the DistilRoBERTa-financial-sentiment model to classify individual financial news headlines as positive, negative, or neutral. The model output is aggregated into a daily normalized sentiment score $\text{sent}_t \in [0,100]$, which, along with the VIX, forms the sentiment input to the agents.

By synthesizing the Forecasting Agent's return forecast $f_t$ with the daily sentiment score $\text{sent}_t$, the system effectively constructs a continuous market-regime indicator, allowing both Trading and Hedging agents to recognize and adapt to shifts between benign and stressed market environments.

**Joint Learning and Agent Coordination**

To ensure consistent decision-making across agents, we adopt a shared reward structure based on the Sharpe ratio of the overall portfolio (Sharpe, 1994). At each time step $t$, both the Trading and Hedging Agents receive the same scalar reward:

$$R_t = SR_t - SR_{t-1},$$

where $SR_t = \frac{E[r_t] - r_f}{\sigma(r_t)}$ with $r_f$ as the risk-free rate and $\sigma(r_t)$ the standard deviation of returns. This formulation encourages improvements in risk-adjusted performance rather than raw returns alone.

In addition, we incorporate a lightweight inter-agent communication mechanism using cross-attention. Each agent produces a compact internal summary that reflects its current decision context—such as its interpretation of the market and its intended course of action. This summary is shared with the other agent through a cross-attention mechanism, which generates a context vector that is appended to the receiving agent's observation. By exchanging this high-level information about each other's intentions, the agents can adapt their strategies in a coordinated and mutually informed manner, leading to improved system-wide robustness.

**Trading Agent**

The Trading Agent aims to maximize portfolio returns using a PPO-based reinforcement learning framework that determines optimal stock trading actions. It extends the global state by including a cross-attentional context vector $c_t^{hedge}$, representing latent signals received from the Hedging Agent. The agent selects an action $a_t \in [-1, 1]$, representing a continuous allocation decision. A value of $a_t = -1$ corresponds to full liquidation of the equity position, $a_t = 1$ indicates full investment of all available cash into the asset, and $a_t = 0$ means maintaining the current holdings. Intermediate values result in proportional adjustments—for example, $a_t = 0.4$ allocates 40% of available cash.

The desired number of shares $N_t^{des}$ is defined by:

$$N_t^{des} = \begin{cases} N_{t-1} + \alpha_t \dfrac{b_{t-1}}{p_t}, \alpha_t > 0 \\ N_{t-1}(1 + \alpha_t), \alpha_t < 0 \end{cases}.$$

Let $N_{t-1}$ be the number of shares held before trading, then the number of shares traded is:

$$\Delta N_t = N_t^{des} - N_{t-1}.$$

To account for trading expenses, we adopt a transaction cost model commonly used in prior research, given by: $C_t = 0.002 \times p_t \times |\Delta N_t|$. Rather than computing a private profit-minus-cost reward, the Trading Agent receives the shared portfolio-level reward $R_t$ defined earlier, ensuring its updates align directly with improvements in the overall Sharpe ratio.

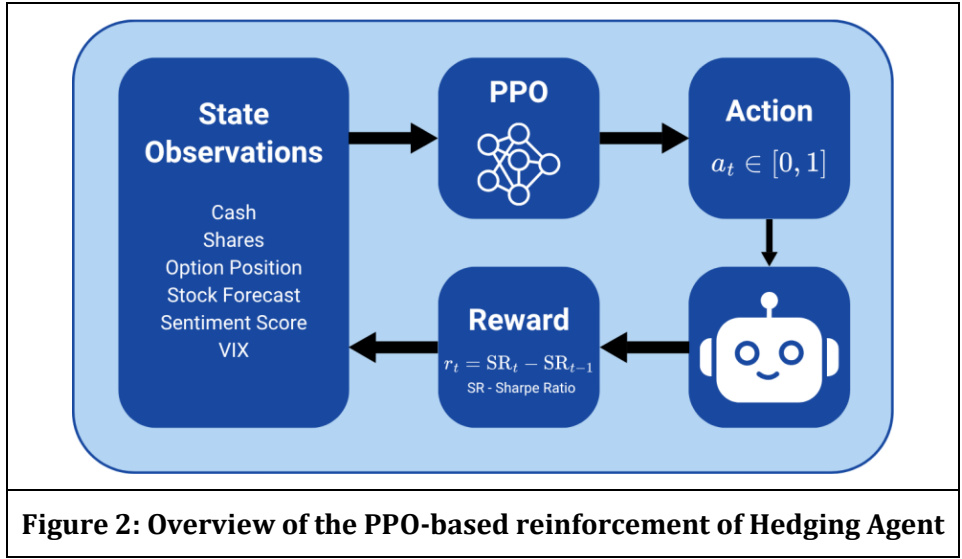

**Figure 2: Overview of the PPO-based reinforcement of Hedging Agent**

### Hedging Agent

The Hedging Agent dynamically adjusts the portfolio's protective option positions to mitigate downside risk, striving to maintain a delta-neutral portfolio in response to evolving market conditions (Hull, 2006). It extends the global state by including a context vector $c_t^{trade}$, which reflects the Trading Agent's hidden intent at time $t$. This enables the Hedging Agent to align its protection decisions with the trading strategy in a coordinated manner. Each share of the underlying stock has a positive delta of $+1$, while a put option has a negative delta, $\Delta_{put} < 0$. A strictly delta-neutral position requires that the sum of deltas from the stock and the put options be zero. In other words, if $h_t$ is the number of shares, then the full delta-hedge would be: $n_t = -\frac{h_t}{\Delta_{put}}$. However, to allow for partial hedging, the Hedging Agent chooses a continuous hedge ratio $\alpha_t \in [0,1]$, which determines the fraction of a full delta-hedge to apply. Thus, the number of puts traded at time $t$ becomes:

$$n_t = \alpha_t \cdot \left(-\frac{h_t}{\Delta_{put}}\right),$$

where $\alpha_t = 1$ implies a fully delta-neutral stance, and $\alpha_t = 0$ implies no hedging. If each option contract covers multiple shares (e.g., 100), you would scale accordingly.

Option trades are priced at executable levels using bid/ask quotes—purchases at the ask, sales at the bid. The total hedging cost is modeled as:

$$C_t^{hedge} = n_t \cdot 0.007 + 0.005 \cdot n_t \cdot P_t^{opt},$$

where 0.007 is the fixed cost per contract (e.g., \$0.70), $P_t^{opt}$ is the option premium per contract (ask price if buying, bid if selling), and 0.005 reflects a 0.5% proportional cost relative to the premium (Yan, 2022). This formulation captures both execution slippage and brokerage fees, and varies realistically with market conditions and order size. Execution is permitted only when market volume is sufficient to support the trade size $n_t$; if the available open interest or volume is below the required threshold, the order is deferred or scaled down proportionally.

Like the Trading Agent, the Hedging Agent receives the shared portfolio-level reward $R_t = SR_t - SR_{t-1}$ defined earlier, which aligns both agents toward improving the overall Sharpe ratio of the portfolio. Over repeated daily interactions, the Hedging Agent observes its state, selects a hedge ratio $\alpha_t$, and updates its policy to maximize $R_t$. We employ a reinforcement learning mechanism that iteratively refines the agent's decisions based on feedback from market outcomes (Z. Li et al., 2024; X.-Y. Liu et al., 2022). *Figure 2* illustrates this learning process, showing how state observations guide the agent's actions and reward function. By incrementally improving $SR_t$ relative to $SR_{t-1}$, the Hedging Agent balances potential drawdown protection against the costs of purchasing options, ultimately enhancing risk-adjusted returns in complex market environments.

**Coordinator Agent**

The Coordinator Agent synchronizes the workflow and updates the overall portfolio state. Each trading day, it:

1. **Retrieves Data:** Collects updated asset prices and option quotes.
2. **Checks Expired Options:** At each time ($t$), the Coordinator verifies whether any options have expired. For an option with expiration ($t_{\exp}$), update:
$$b_{\{t+1\}} = b_t + \mathbf{1}_{\{t=t_{exp}\}} \cdot n_t \cdot M \cdot max(0, K - p_t)$$
$$n_{\{t+1\}} = n_t \cdot \mathbf{1}_{\{t \neq t_{exp}\}}$$
   where ($K$) is the strike price and ($\mathbf{1}\{\cdot\}$) is the indicator function.
3. **Generates Signals:** Invokes the Forecasting and Sentiment Analysis Agents to update ($f_t$) and (sent$_t$).
4. **Executes Trading:** Applies the Trading Agent's chosen action ($a_t$) to update ($h_t$) and ($b_t$).
5. **Executes Hedging:** Applies the Hedging Agent's decision to adjust option positions.
6. **Computes Reward:** After both trades are applied, the Coordinator computes the shared portfolio-level reward $R_t$ and passes it to both agents for policy updates. It also stores the internal decision summaries of each agent so they can be used as input context for the next time step's cross-attention mechanism.
7. **Updates State:** Finally, the Coordinator updates the state for ($t + 1$):
$$s_{t+1} = [p_{t+1}, h_{t+1}, b_{t+1}, n_{t+1}, O_{t+1}, f_{t+1}, \text{sent}_{t+1}, \text{VIX}_{t+1}].$$

The overall flow of information between the Coordinator Agent and specialized agents in the system is depicted in Figure 3, illustrating how trading signals, sentiment analysis, and hedging strategies are processed to make portfolio adjustments.

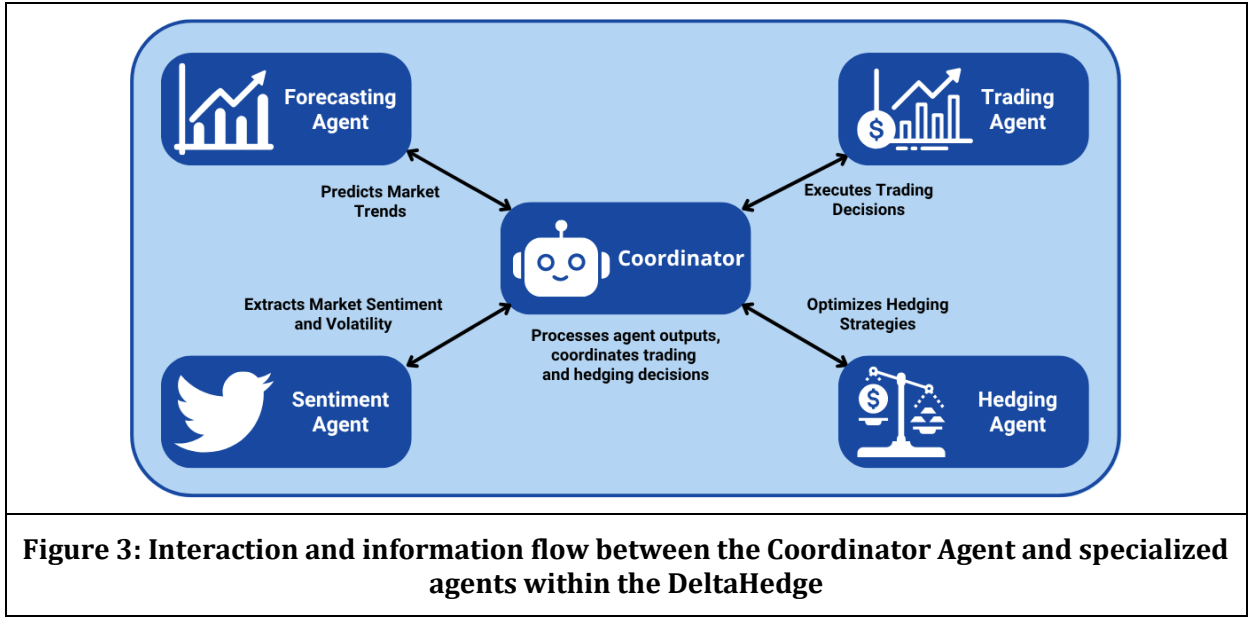

**Figure 3: Interaction and information flow between the Coordinator Agent and specialized agents within the DeltaHedge**

*Ensemble Strategy for Hedging*

Given the dynamic nature of financial markets, different RL algorithms may excel under varying conditions. To enhance robustness, we adopt an ensemble strategy (Szydlowski & Chudziak, 2024; Yang et al., 2020)— training multiple candidate Hedging Agents (PPO, DDPG, and A2C) concurrently, then choosing the best performer for live hedging.

**Candidate Policies**

Let the trained hedging policies be $\{\pi_1, \pi_2, \dots, \pi_N\}$, where each $\pi_i$ is learned by a distinct RL algorithm (PPO, A2C, or DDPG). In principle, each agent aims to hedge market exposure by producing an action $a_{t,i}$ given the current state $s_t$. The hedging reward associated with agent $i$ at time $t$ is denoted $R_{t,i}$.

**Concurrent Training and Validation**

- **Concurrent Training:** Every quarter, we train each policy $\pi_i$ in parallel on recent historical data. This rolling retraining is motivated by empirical studies showing that market dynamics can shift significantly over multi-month horizons (Yang et al., 2020).
- **Validation:** Following training, we reserve a one-month validation window to measure each agent's risk-adjusted performance. Let $\mathcal{M}(\pi_i)$ denote this performance metric—often the Sharpe Ratio:

$$\mathcal{M}(\pi_i) = \frac{E[r(\pi_i)]}{\text{StdDev}[r(\pi_i)]}$$

  computed from the daily returns $r(\pi_i)$ under policy $\pi_i$ in the validation set.

- **Selection:** We activate only the best agent $i^*$ for live hedging:

$$i^* = \arg \max_{i \in \{1, \dots, N\}} \mathcal{M}(\pi_i),$$

  meaning $\pi_{i^*}$ is used to generate hedging actions until the next retraining cycle. In other words, our deployed ensemble policy is simply:

$$\pi^*(a_t \mid s_t) = \pi_{i^*}(a_t \mid s_t), \quad R_t = R_{t,i^*}.$$

**System Objective**

Although multiple policies $(\pi_1, \dots, \pi_N)$ are trained in parallel, only the top-performing policy $(\pi_{i^*})$ is deployed at any time. All candidate agents are trained using the same shared reward signal $R_t$, based on changes in the portfolio's Sharpe ratio. The learning objective for the selected policy is then denoted:

$$J(\theta_{i^*}) = E_{\pi_{i^*}}[R_t],$$

where $\theta_{i^*}$ are the parameters of the chosen agent and $R_t$ reflects the improvement in risk-adjusted return. Internally, each RL algorithm (e.g., PPO, A2C, DDPG) may apply its own clipped or regularized version of this objective, but all agents are aligned with the same high-level portfolio goal.

**Quarterly Retraining Rationale**

Periodic (quarterly) retraining strikes a balance between adaptability—ensuring the hedging policy remains attuned to recent market conditions—and computational efficiency—avoiding overly frequent training. In each cycle:

1. **Lookback window** (90 days) for training all $\pi_i$.
2. **Short validation window** (30 days) to compute Sharpe Ratios $\mathcal{M}(\pi_i)$.
3. **Activate** best agent $\pi_{i^*}$ until the next retraining trigger.

By rotating among multiple RL algorithms and selecting the top performer, we reduce the risk of relying on a single model that may underperform in specific market environments. Consequently, our ensemble-like approach aims to maximize portfolio value over time, delivering robustness in the face of shifting market regimes.

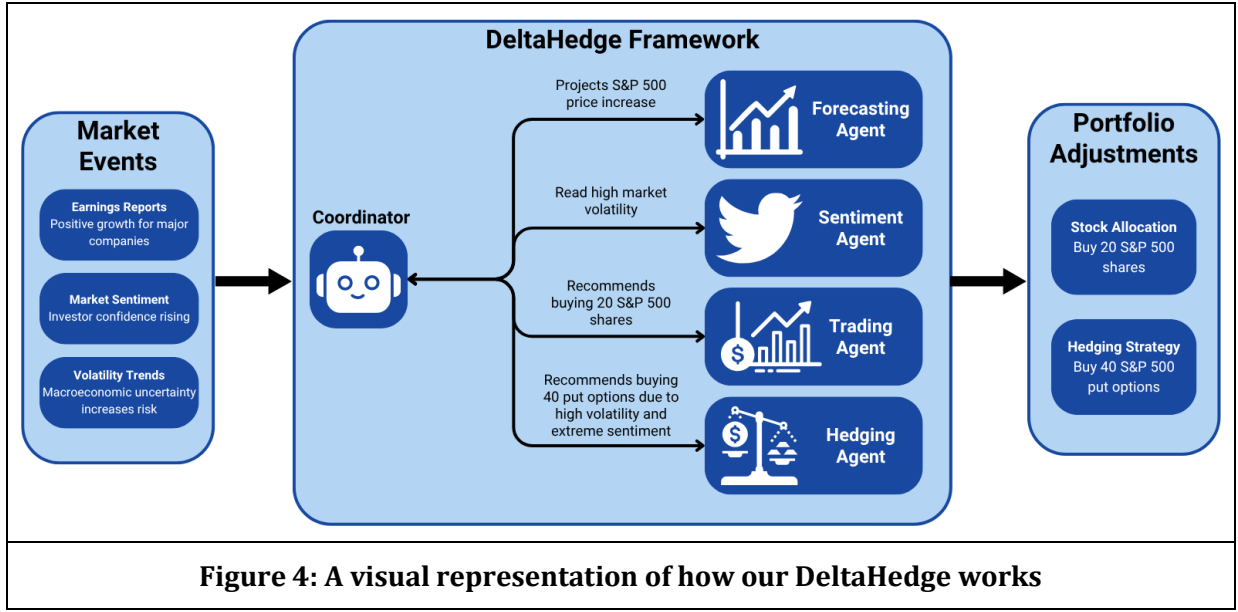

**Figure 4: A visual representation of how our DeltaHedge works**

### Practical Use Case: Navigating a Moderately Bullish Yet Volatile Market

To illustrate how the system reacts under a moderately bullish yet volatile market, we consider a scenario in which S&P 500 forecasts turn positive (e.g., +2% price increase) while daily sentiment data indicates rising investor confidence. Despite this optimistic outlook, macroeconomic uncertainty remains high, prompting the Hedging Agent to recommend purchasing near-the-money SPX put options covering a fraction of the newly acquired equity stake (e.g., 20 SPX shares). By balancing the Trading Agent's bullish equity allocation with protective puts, the system aims to stabilize daily changes in the portfolio's Sharpe Ratio and prevent severe drawdowns if market conditions unexpectedly deteriorate.

As depicted in Figure 4, the Coordinator Agent consolidates signals from the Forecasting and Sentiment Agents to inform the Trading Agent's equity purchases. Concurrently, it applies the Hedging Agent's recommendation to acquire protective puts, thereby mitigating potential losses if market sentiment or economic indicators worsen (Pang et al., 2022). By promptly adjusting the hedge ratio whenever predictive signals shift, the system maintains a more stable risk profile and avoids large drawdowns.

## Experiments and Results

In this section, we evaluate the performance of our multi-agent framework. We compare our approach against traditional financial strategies, standalone RL models, and alternative variations of our framework, including single-agent hedging and trading without options integration. The goal is to determine whether our ensemble-based hedging strategy enhances risk-adjusted returns and portfolio stability.

We first introduce the experimental setup, including the datasets and simulation environment. Next, we conduct benchmark comparisons against multiple baselines, analyzing their performance through Sharpe Ratio, Sortino Ratio, maximum drawdown, annualized returns, and portfolio volatility. Finally, we discuss the implications of our findings and the impact of options integration.

### Experimental Setup

The experiments are conducted in a simulated trading environment designed to emulate real-world market conditions. The environment incorporates historical prices, options data, and sentiment signals extracted from financial news and social media. Our framework operates on a daily time scale, processing new market data at each step and adjusting the portfolio accordingly. We evaluate the system across three assets: the S&P 500 index, Apple (AAPL), and Tesla (TSLA), to test its robustness across different market dynamics.

**Datasets**

The following datasets are used:

- **Stock price data**: Daily closing prices and historical volatility from Yahoo Finance.

- **Options data**: Retrieved from optionsdx.com, including key contract information used for hedging decisions.

- **Sentiment data**: News headlines and content retrieved from EOD Historical Data, used to compute sentiment scores via a transformer-based classifier.

- **Volatility index (VIX)**: Sourced from Yahoo Finance to capture overall market uncertainty.

The training period spans from January 2010 to December 2019 for S&P 500 and from January 2016 to December 2019 for both Apple and Tesla stocks. During this phase, all models are trained using historical data. Additionally, the Hedging Agents are retrained throughout the testing phase (January 2020 to December 2024) using a rolling-window strategy. Every 90 trading days, new candidate policies are trained on recent data, and after a 30-day validation window, the best-performing policy is selected for deployment. This approach allows the system to remain adaptive during evaluation while ensuring that all model decisions are based strictly on past information, thereby preventing data leakage.

**Model Training and Implementation**

In our experiments, each reinforcement learning model (PPO, DQN, A2C, and DDPG) was implemented using Stable-Baselines3 and trained for 20,000 timesteps. Although algorithm-specific hyperparameters (such as exploration strategies or advantage clipping) vary, we applied a broadly consistent configuration across all methods. In addition, each model employed the same Informer-based feature extractor, ensuring that temporal dependencies were captured uniformly. This setup provides a fair comparison of the different algorithms while minimizing discrepancies due to implementation details.

## *Benchmark Comparisons*

We compare our framework against three key baselines:

1. **Traditional Strategies**

   o We evaluate **Buy-and-Hold (B&H)**, and **KDJ with an RSI filter**, representing common technical trading approaches. These methods rely on pre-defined rules and do not adapt dynamically to changing market conditions.

2. **Standalone RL Models**

   o We test **PPO** and **DQN** as standalone RL trading agents. Unlike our framework, these models focus solely on trading the underlying asset, without options-based risk management.

3. **LLM-Based Frameworks**

   o We include **FinAgent**, a large-language-model–driven trading system that generates daily allocation decisions from financial prompts.

The primary objective is to assess whether our **ensemble strategy outperforms** static rules, equity-only RL (with and without a hand-crafted hedge), and modern LLM-driven trading.

## *Ablation Studies*

To isolate the contributions of our architecture's components, we run three in-house variants:

1. **Single Hedging Agent (Unensembled)**

   o We compare our ensemble strategy against three separate trials, where a single RL model (PPO, A2C, or DDPG) handles hedging. This allows us to measure whether switching between multiple models provides a tangible performance benefit.

2. **Without Options Integration**

   o We assess the performance of our full multi-agent system without the hedging agent, meaning the sentiment, forecasting, and trading agents operate alone. This comparison isolates the impact of hedging on overall portfolio stability.

3. **Classical Delta-Hedge**

   o A hand-crafted benchmark that applies a full delta-neutral put hedge whenever the continuous regime indicator turns bearish. Allows us to compare our learned hedgers against textbook practice.

By comparing our full ensemble strategy against the Single-Hedger and No-Options variants, we isolate exactly how dynamic, coordinated options hedging drives improvements in performance metrics.

## *Performance Metrics and Results*

We assess each strategy using six key metrics:

- **Sharpe Ratio (SR)**: Measures risk-adjusted return by dividing excess returns by volatility.

- **Sortino Ratio (SoR)**: A variation of the Sharpe Ratio that penalizes downside volatility.

- **Calmar Ratio (CR)**: The ratio of annualized return to maximum drawdown, showing how much return is earned per unit of drawdown.

- **Total Return (TR)**: The cumulative percentage return over the testing period, reflecting overall portfolio performance.

- **Maximum Drawdown (MDD)**: The worst peak-to-trough decline, indicating risk exposure.

- **Volatility (Vol)**: The standard deviation of returns, reflecting price fluctuations.

In addition to the overall backtest, we also evaluate each strategy over three distinct market regimes for S&P 500—Rapidly Rising (April 1, 2020 – August 31, 2021), Rapidly Falling (January 1, 2022 – June 30, 2022) and Extremely Volatile (May 1, 2022 – January 31, 2023)—to assess how performance varies under different volatility environments. We believe that employing this set of metrics and tests provides a comprehensive view of how well a strategy manages risk and generates returns in varied market conditions. We also present visual analytics to further illustrate differences in performance and offer deeper insights.

## *Results Overview*

Table 1 provides a detailed comparison of all benchmark strategies on the S&P 500. DeltaHedge achieves a Sharpe ratio of 1.33 and a Sortino ratio of 1.81—figures that nearly double those of the next-best method—while delivering a total return of 121 % with a maximum draw-down of only 10 %. Its annualized volatility is just 14 %, lower than any other algorithm except the classic delta hedge, which sacrifices a quarter of the return to achieve similar risk. Figure 5 confirms that DeltaHedge's equity curve consistently overtakes

| Table 1. Performance Comparison Table of All Baselines for S&P 500 | | | | | | |
|---|---|---|---|---|---|
| **Strategy** | **SR** | **SoR** | **CR** | **TR(%)** | **MDD(%)** | **Vol(%)** |
| KDJ (RSI Filter) | 0.25 | 0.32 | 0.14 | 15.84 | 27.66 | <u>18.33</u> |
| Buy-and-Hold (B&H) | 0.44 | 0.55 | 0.30 | 45.36 | 33.19 | 22.49 |
| Standalone PPO | 0.49 | 0.61 | 0.34 | 49.56 | 31.55 | 21.38 |
| Standalone DQN | 0.53 | 0.66 | 0.39 | 53.16 | 29.03 | 20.51 |
| FinAgent | <u>0.81</u> | <u>1.04</u> | <u>0.86</u> | <u>90.42</u> | <u>20.36</u> | 19.93 |
| **DeltaHedge** | **1.33** | **1.81** | **2.18** | **121.13** | **10.11** | **14.17** |
| *Improvement (%)* | *64.20* | *74.04* | *153.49* | *33.94* | *-50.34* | *28.88* |

the others after each market sell-off. What is not immediately obvious is that the strategy's option layer does more than smooth linear price swings: by dynamically reshaping the payoff it monetizes tail events, boosting both return and downside protection. Indeed, compared with FinAgent's 90 % return and 20 % draw-down, DeltaHedge adds roughly one-third more gain while halving the worst loss. We ran standard bootstrap tests on mean excess returns and on Sharpe ratios, and in both cases the p-values fall below 0.10, confirming that these improvements are statistically significant at the ten-percent level.

| Table 2. Performance Comparison Table of All Baselines for Apple and Tesla | | | | | | | | |
|---|---|---|---|---|---|---|---|---|
| **Strategy** | **Apple** | | | | **Tesla** | | | |
| | SR | SoR | TR% | MDD% | SR | SoR | TR% | MDD% |
| KDJ (RSI Filter) | 0.49 | 0.69 | 50.14 | <u>25.88</u> | 1.06 | 1.40 | 293.06 | 64.54 |
| Buy-and-Hold (B&H) | 0.83 | 1.18 | 163.16 | 31.42 | 1.11 | 1.63 | 766.21 | 73.63 |
| Standalone PPO | 0.84 | 1.19 | 164.60 | 31.42 | 1.18 | 1.83 | 410.73 | 51.80 |
| Standalone DQN | 0.88 | 1.20 | 176.45 | 30.94 | 1.20 | 1.77 | 568.32 | 51.80 |
| FinAgent | <u>0.99</u> | <u>1.35</u> | <u>188.46</u> | 30.03 | <u>1.58</u> | <u>2.29</u> | <u>662.88</u> | <u>40.35</u> |
| **DeltaHedge** | **1.09** | **2.37** | **254.87** | **13.15** | **1.71** | **2.55** | **984.52** | **26.95** |
| *Improvement (%)* | *10.01* | *75.56* | *35.23* | *-49.17* | *8.23* | *11.35* | *28.48* | *33.23* |

Turning to single stocks, Table 2 shows that the same hierarchy holds for Apple and Tesla. On Apple, DeltaHedge records a Sharpe of 1.09 and limits draw-down to 13 %, roughly half of FinAgent's; on the far more volatile Tesla, it compounds nearly 1 000 % while capping draw-down at 27 %. Figure 5 illustrates how the ensemble's curve separates decisively after each 2022–23 whipsaw. Interestingly, buy-and-hold already returns 766 % on Tesla, yet DeltaHedge augments that by another 220 % and roughly halves the maximum loss—evidence that the framework truly "harvests" volatility rather than merely insures against it. The smaller gap between DeltaHedge and FinAgent on Apple suggests that deep hedging is most valuable when tail moves become extreme.

| Table 3. Ablation Study: Impact of Hedging Variants for S&P 500 | | | | | | |
|---|---|---|---|---|---|---|
| **Strategy** | **SR** | **SoR** | **CR** | **TR(%)** | **MDD(%)** | **Vol(%)** |
| Without Hedging Agent | 0.64 | 0.85 | 0.66 | 64.87 | 20.36 | 19.32 |
| Classic Delta Hedging | 0.95 | 1.34 | 1.00 | 76.42 | 15.22 | 13.92 |
| Single Hedging (A2C) | 1.15 | 1.53 | 1.84 | 96.89 | <u>10.05</u> | <u>13.89</u> |
| Single Hedging (PPO) | <u>1.17</u> | <u>1.55</u> | <u>1.89</u> | 98.95 | **9.94** | **13.88** |
| Single Hedging (DDPG) | <u>1.17</u> | <u>1.55</u> | 1.86 | <u>99.53</u> | 10.17 | <u>13.89</u> |
| **DeltaHedge** | **1.33** | **1.81** | **2.18** | **121.13** | 10.11 | 14.17 |
| *Improvement (%)* | *13.68* | *16.77* | *15.34* | *21.58* | *–* | *–* |

The ablation study on the S&P 500 (Table 3 and Figure 6) confirms that every hedging layer adds value over the unhedged baseline, yet the full DeltaHedge ensemble still delivers the strongest risk–return trade-off. The best standalone agent (PPO) achieves a Sharpe of 1.17, while the ensemble boosts it to 1.33 and raises the Calmar ratio above 2 for the first time. Classic delta hedging does reduce draw-down from 20 % to 15 %, but at the expense of surrendering over twenty percentage points of total return; by contrast, the learning-based hedgers confine draw-downs to roughly 10 % without giving up upside. Notably, A2C, PPO and DDPG all trace almost identical return paths—with only a tenth of a volatility point between them—suggesting that simply combining models is not enough. Instead, the quarterly policy-switching mechanism, which retires underperforming specialists and promotes those best suited to the current regime, appears to drive the ensemble's extra lift. Moreover, the ensemble's annualized volatility of 14.2 % is only marginally higher than that of any single agent, meaning that the additional return actually comes at negligible incremental risk. This pattern of stable volatility alongside rising returns underscores that the dynamic adaptation, rather than any one model, underpins DeltaHedge's superior performance.

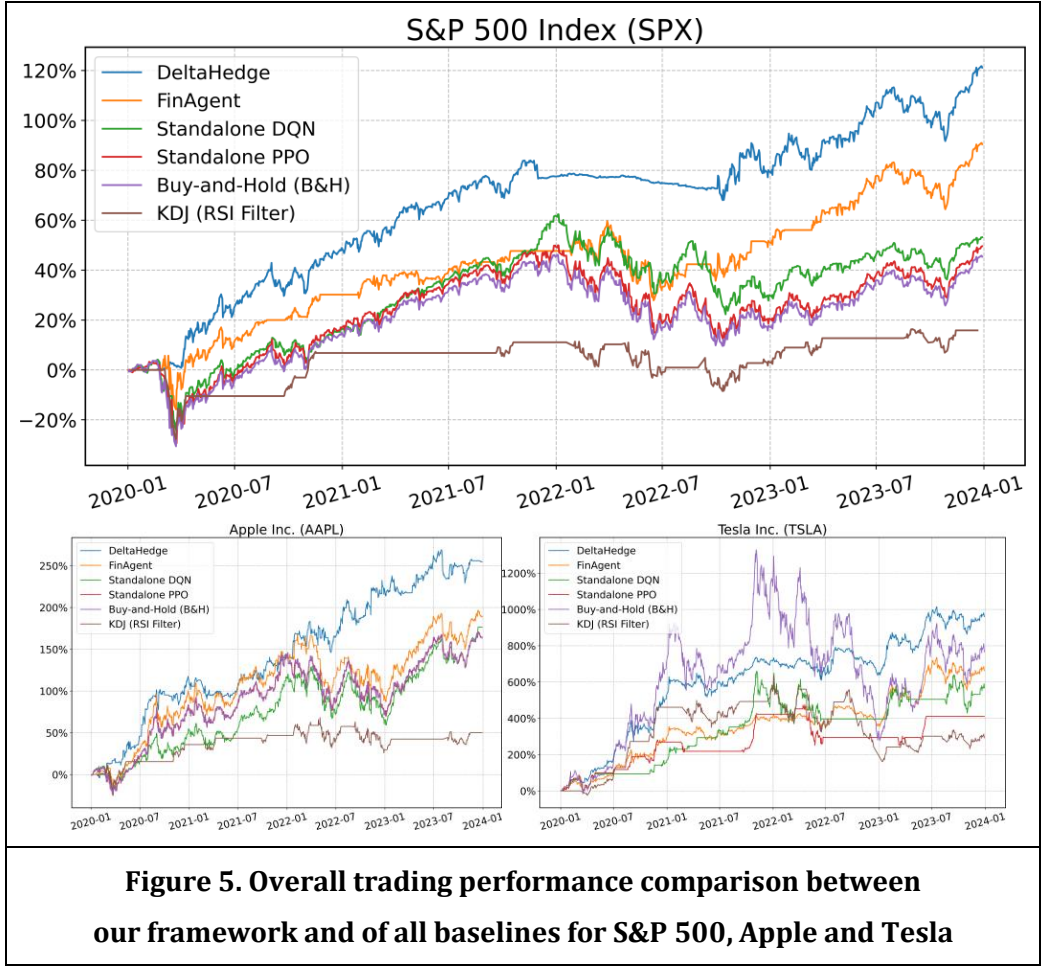

**Figure 5. Overall trading performance comparison between our framework and of all baselines for S&P 500, Apple and Tesla**

On Apple and Tesla, the ablation results in Table 4 follow a similar pattern. For Apple, the ensemble outperforms the best single hedger by only five basis points in Sharpe but halves the draw-down to 13 %; for Tesla, the Sharpe advantage is similarly small, yet draw-down falls by three percentage points. Notably, the classic delta hedge is second-best on Tesla but worst on Apple. A fixed delta rule suffices when an underlying has high gamma, yet it underreacts to smoother trends in a mega-cap stock. Across both names the ensemble never posts the highest raw return, but it always delivers the smallest draw-down—evidence that the switching logic implicitly prioritizes downside control over return maximization.

| | **Apple** | | | | **Tesla** | | | |
|---|---|---|---|---|---|---|---|---|
| **Strategy** | SR | SoR | TR% | MDD% | SR | SoR | TR% | MDD% |
| Without Hedging Agent | 1.03 | 1.45 | 160.08 | <u>23.41</u> | 1.13 | 1.58 | 726.90 | 67.67 |
| Classic Delta Hedging | 0.75 | 0.99 | 122.05 | 26.66 | 1.67 | 2.55 | 837.51 | 27.51 |
| Single Hedging (A2C) | 0.99 | 1.36 | 189.17 | 30.04 | <u>1.90</u> | <u>2.90</u> | 942.08 | **23.00** |
| Single Hedging (PPO) | <u>1.04</u> | <u>1.47</u> | 214.68 | 30.11 | 1.71 | 2.55 | **984.52** | 26.95 |
| Single Hedging (DDPG) | <u>1.04</u> | <u>1.47</u> | <u>215.56</u> | 30.09 | 1.85 | 2.83 | 976.89 | 24.39 |
| **DeltaHedge** | **1.09** | **2.37** | **254.87** | **13.15** | **1.92** | **2.91** | 960.66 | <u>23.07</u> |
| *Improvement (%)* | *4.81* | *61.22* | *18.25* | *-43.82* | *1.05* | *0.34* | *-* | *-* |

**Table 4. Ablation Study: Impact of Hedging Variants for Apple and Tesla**

Examining Figures 5 and 6, we observe that strategy performance trajectories exhibit distinct "signature patterns" during market transitions. DeltaHedge consistently demonstrates smoother curve gradients during volatility spikes, while other strategies show jagged, reactive movements. This visual smoothness correlates with lower drawdowns but does not sacrifice responsiveness to genuine trend changes. Particularly noteworthy is the visual divergence between benchmark and ablation plots for Tesla, where the extreme price movements create visual "compression zones" that mask significant performance differences until they suddenly manifest as dramatic separations in the curves. The temporal clustering of these separation events suggests that strategy differentiation occurs primarily during specific market catalyst moments rather than gradually over time. This visual insight complements the quantitative analysis by highlighting the episodic nature of performance advantages, which has important implications for the practical implementation of trading strategies in real-world market environments.

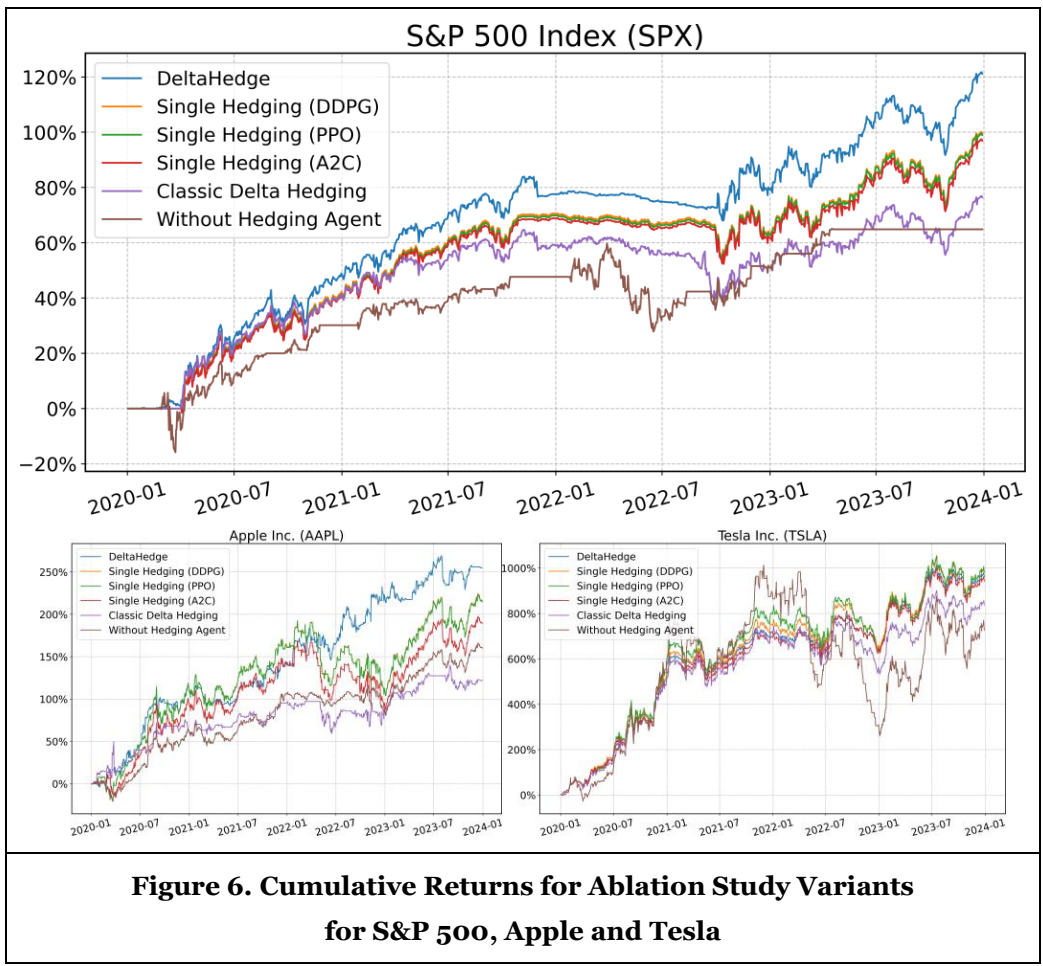

**Figure 6. Cumulative Returns for Ablation Study Variants**

**for S&P 500, Apple and Tesla**

Figure 7 breaks the sample into three market regimes: rapid rallies, sharp sell-offs and highly volatile sideways moves. In rising markets, DeltaHedge keeps pace with and often exceeds buy-and-hold despite the cost of financing puts. During the Covid crash of 2020 and the 2022 inflation-driven decline, the ensemble prevents any breach below the previous cycle low, whereas every benchmark suffers a new trough. In the whipsaw conditions of 2023, single-model hedgers churn through option premium and flatten out, but the ensemble throttles back protection just enough to preserve convexity for the subsequent rebound. Taken together, these regime-based observations show that the ensemble behaves like an adaptive long-gamma position: it lightens hedges when insurance is overpriced, but it switches instantly to full protection when volatility materializes.

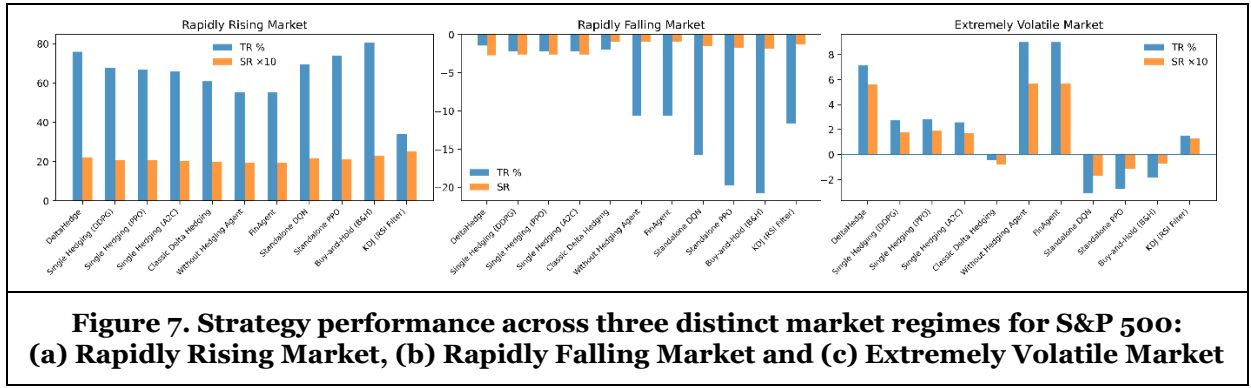

**Figure 7. Strategy performance across three distinct market regimes for S&P 500: (a) Rapidly Rising Market, (b) Rapidly Falling Market and (c) Extremely Volatile Market**

Stepping back from the numbers, several broader insights emerge. First, the option market rewards agents that can time protection rather than buying insurance indiscriminately. Second, reinforcement learning can uncover such timing rules from high-frequency price data alone, without relying on explicit macro inputs. Third, the performance gains come not from finding a single "perfect" model but from maintaining a pool of candidates and selecting the one best suited to current conditions. In other words, the strength of the ensemble lies in its flexibility: when volatility spikes, a more conservative hedger takes precedence, but during calm periods, a more aggressive policy captures upside. In practical terms, this means that embracing multiple strategies and allowing live performance to determine which policy is active can be far more effective than spending time searching for a single ideal model.

## Conclusion and Future Work

In this paper, we introduced **DeltaHedge**, a hierarchical multi-agent framework that leverages options trading and reinforcement learning to enhance both portfolio performance and downside protection. By deploying specialized agents for forecasting, sentiment analysis, trading, and hedging—coupled with an ensemble-based mechanism that selects the most effective RL hedging policy at any given time—our approach effectively adapts to changing market conditions and mitigates risk.

Empirical results on the S&P 500, Apple and Tesla demonstrate that DeltaHedge more than doubles risk-adjusted metrics relative to standard benchmarks (Sharpe ratios up to 1.33 and Sortino ratios up to 1.81) while capping draw-downs at one third of those seen under buy-and-hold or pure RL methods (Table 1 and Table 2; Figure 5). Ablation studies confirm that each layer of hedging contributes to overall gains—but only the full ensemble delivers the best blend of return and risk control.

Our work contributes to the field of multi-agent systems for finance by merging cutting-edge reinforcement learning techniques with proven options-based hedging methods, thereby bridging the gap between theoretical models and practical portfolio management. This approach not only reinforces the critical role of options in balancing risk and return but also highlights the benefits of adaptive, multi-agent coordination in complex market environments.

There are several promising avenues to extend this research. First, we plan to enrich the hedging toolkit by incorporating more advanced mathematical strategies (Escobar-Anel et al., 2022)—such as collars, straddles and other multi-leg option structures—allowing the agents to tailor their protection more precisely to evolving risk profiles. Second, we aim to generalize the framework from single-asset to multi-asset portfolios, which would enable sector-level allocation, cross-instrument hedging and broader diversification benefits (Shu et al., 2024). Finally, we will explore richer inter-agent interaction schemes—such as structured debate or peer-review protocols—so that forecasting, trading and hedging agents can challenge one another's proposals before execution, fostering a kind of internal "markets of ideas" that may yield more robust, consensus-driven decisions.